\documentclass[12pt]{iopart}
\usepackage{iopams} 
\usepackage{graphics}
\usepackage{graphicx}
 
\begin{document}

\title{High error-rate quantum key distribution for long-distance communication}
\author{Muhammad Mubashir Khan}
\address{The School of Computing, University of Leeds, Leeds, LS2 9JT, United Kingdom}
\ead{mmkhan@comp.leeds.ac.uk}
\author{Michael Murphy}
\address{Institut f\"{u}r Quanteninformationsverarbeitung, Universit\"{a}t Ulm, Albert-Einstein-Allee 11, 89081 Ulm, Germany}
\author{Almut Beige}
\address{The School of Physics and Astronomy, University of Leeds, Leeds, LS2 9JT, United Kingdom}

\begin{abstract}
In the original BB84 protocol by Bennett and Brassard, an eavesdropper is detected because his attempts to intercept information result in a quantum bit error rate (QBER) of at least $25 \, \%$. Here we design an alternative quantum key distribution protocol, where Alice and Bob use two mutually unbiased bases with one of them encoding a ``0" and the other one encoding a ``1." The security of the scheme is due to a minimum index transmission error rate (ITER) introduced by an eavesdropper which increases significantly for higher-dimensional photon states. This allows for more noise in the transmission line, thereby increasing the possible distance between Alice and Bob without the need for intermediate nodes.
\end{abstract}

\maketitle 

\section{Introduction} \label{intro}

The aim of quantum cryptography is to establish a shared, secret, and random sequence of bits between a sender, called {\em Alice}, and a receiver, called {\em Bob} \cite{review}. This sequence constitutes a perfect cryptographic key, a so-called one-time pad, and allows Alice and Bob to securely encrypt a message of the same length. The cryptographic key is obtained solely via the transmission of photons and classical communication. Each bit is encoded in the state of a single photon and read out by Bob upon arrival via a quantum measurement. Random switching between different bases makes it impossible for an eavesdropper, called {\em Evan}, to predict the states used in the protocol. All his attempts to intercept photons result in a significant {\em quantum bit error rate} (QBER). This guarantees a high level of security, since Evan's presence is detected easily when Alice and Bob compare a number of test bits.

Under ideal conditions, the exchange of single photons allows Alice and Bob to establish a cryptographic key over an arbitrarily long distance. In practice, cryptographic setups consist of imperfect single photon sources, lossy transmission lines, and photon detectors with dark count rates. Alice and Bob must hence apply classical information processing tools like error correction and privacy amplification \cite{Deutsch,Ho} to their data in order to obtain identical secret keys. However, cryptographic protocols are only secure as long as it is possible to detect the presence of an eavesdropper. System errors could cloud Evan's presence; especially, since he could simply replace parts of the equipment with high quality components. This makes it impossible to tolerate large system errors and limits the possible distance between Alice and Bob.  

Recently, D. Rosenberg {\em et al.} \cite{losalamos,losalamos2} reported the creation of a secure cryptographic key over a distance of $144.3\,$km of optical fiber. Their scheme is based on a decoy state protocol \cite{decoy1,decoy2,decoy3} which is immune to photon number splitting attacks and highly resistant to Trojan horse attacks \cite{Kraus}. It is expected that improvements in filtering of blackbody photons might allow for an extension of the fiber to $250 \,$km. In the mean time, H. Takesue {\em et al.} \cite{yamamoto} and D. Stucki {\em et al.} \cite{hugo} created a secure cryptographic key over a distance of $200 \,$km of optical fiber. These experimental setups are believed to be the longest terrestrial quantum key distribution fiber-links yet demonstrated. Comparable distances have been achieved in free space. For example, T. Schmitt-Manderbach {\em et al.} \cite{europe} securly distributed a cryptographic key over a $144\,$km free-space link. 

In this paper, we design a novel quantum key distribution protocol whose minimum error rate in the case of eavesdropping increases with the dimension of the photon states used by Alice and Bob. In this way, we increase the threshold for tolerable system errors without sacrificing the security of the protocol and hence increase the possible distance between Alice and Bob. In principle, single photons could be purified with the help of quantum repeaters \cite{Briegel}. Proposals for their implementation (see e.g.~Refs.~\cite{Lukin,Munro}) and other noise reducing links \cite{Franson,Polzik,Wei} have been made but their experimental implementation and their practical integration into cryptographic networks remains to be seen. 

The above mentioned long-distance quantum key distribution schemes \cite{losalamos,losalamos2,yamamoto,europe} are all based on the BB84 protocol by Bennett and Brassard \cite{BB84}. In BB84, Alice encodes her bits in two-dimensional photon states. These can be obtained using polarisation encoding. However, a more natural choice is time-bin encoding, which affords better protection of the photons against decoherence \cite{timebin}. Alice and Bob independently vary their bases between two possibilities. A key bit is obtained whenever both use the same basis. This means, on average, every second photon contributes a bit to the cryptographic key. Using a simple intercept-resend strategy, an eavesdropper introduces a QBER of at least $25\, \%$ into the communication. 

In the following we assume that Alice and Bob use time-bin or path encoded $N$-dimensional photon states. As in BB84, Alice and Bob randomly vary their bases between two mutually unbiased bases \cite{Wootters}. However, contrary to BB84, Alice and Bob detect the presence of an eavesdropper by calculating the {\em index transmission error rate} (ITER). As we shall see below, for $N=2$, Evan causes a minimum ITER of $25 \, \%$. In case of four-dimensional photon states, this error rate becomes $37.5 \, \%$. When increasing $N$ further, the minimum ITER approaches $50 \, \%$. The efficiency of the protocol in units of transmitted bits per photon is the same as the minimum ITER in case of eavesdropping. The states required by the proposed key distribution scheme can be realised using a symmetric Bell multiport beam splitter \cite{Lim}. Before the transmission, the path encoding of the output states of the Bell multiport beam splitter should be switched to the above mentioned time-bin encoding \cite{timebin}.

Several generalisations of quantum cryptographic schemes to higher dimensions have already been proposed. Refs.~\cite{Bechmann,Bechmann2,Cerf,Cerf2,Bruss,Sych,Walborn} are generalisations of the original BB84 protocol \cite{BB84} based on the encoding of information in higher-dimensional alphabets. However, recently it has been shown \cite{Acin,Scarani,Magniez} that the security of the BB84 protocol is entirely compromised if Alice and Bob share, for example, four-dimensional photon states in this way \cite{Acin2}. Refs.~\cite{highDim,highDim2} propose two alternative quantum cryptographic protocols using four-dimensional photon states which, under ideal conditions, allow Alice and Bob to communicate directly but whose minimum error rates in case of eavesdropping are relatively low. 

Here we show that there are other possible cryptographic schemes in higher dimensions. As in Refs.~\cite{highDim,highDim2}, Alice and Bob use two bases $e$ and $f$ with all states of $e$ encoding a ``0" and all states of $f$ encoding a ``1," even when $N$ is larger than two. This means, contrary to Refs.~\cite{BB84,Bechmann,Bechmann2,Cerf,Cerf2,Bruss,Sych,Walborn}, all vectors of the same basis encode the same bit. Moreover, a bit can be transmitted only when Alice and Bob use {\em different} bases. The quantum key distribution scheme considered in this paper is designed such that no conditions have to be posed on the states of $e$ and $f$, thereby giving us a lot of flexibility when maximising the relevant minimum error rate introduced by Evan. For simplicity, we consider only intercept-resend eavesdropping attacks. This is not the only possible eavesdropping attack, but security against this is considered a strong indication for the general security of a quantum cryptographic protocol. 

In the special case of $N=2$, the cryptographic scheme proposed in this paper is essentially equivalent to the SARG quantum key distribution protocol \cite{hugo2} with the parameter $\chi$ chosen equal to $1/\sqrt{2}$. This protocol is tailored to be robust against photon number splitting attacks. In the SARG protocol, Alice publicly announces which one of the  four different sets of states ${\cal A}_{+,+}$, ${\cal A}_{+,-}$, ${\cal A}_{-,+}$, and ${\cal A}_{-,-}$ she used,  while our protocol requires her only to announce either ``$i=1$" or ``$i=2$.'' This difference is due to a redundancy in the SARG protocol.

There are five sections in this paper. In Section \ref{design}, we introduce the notations used throughout this paper. In Section \ref{eavesdropping}, we calculate the ITER and the QBER for the quantum key distribution protocol introduced in Section \ref{design} as a function of the states used by Alice, Bob, and Evan analytically. Afterwards, we determine their minima in the presence of intercept-resend eavesdropping attacks for different $N$'s  numerically. Geometrical considerations suggest that Alice and Bob should use two mutually unbiased bases. Section \ref{concrete} analyses a concrete protocol based on this idea and shows that mutually unbiased bases indeed guarantee a high ITER and a high QBER in case of eavesdropping. Finally, we summarise our results in Section \ref{conclusions}.

\section{Alternative design} \label{design}

In quantum cryptography there are conventionally three parties, Alice, Bob, and Evan. Alice wants to transmit a sequence of secret bits to Bob. To do so, she prepares single photons in certain states and sends them to Bob. Bob measures the state of each incoming photon. Afterwards, Alice and Bob exchange information via classical communication. At the same time, Evan tries to catch the secret bits without revealing his presence. For example, he measures the state of every transmitted photon and listens in to the classical communication between Alice and Bob. The cryptographic protocol is secure as long as Evan's attempts to obtain information result in an error rate which can be detected easily. 

Let us start by introducing {\em sufficient} conditions for such a protocol to work:
\begin{enumerate}
	\item Bob should measure the incoming photons in a randomly chosen basis. Otherwise, Evan simply uses the same measurement basis and the bit error rate remains zero. This means Bob should randomly switch between {\em at least} two sets of basis states. In the following we assume that this is the case and denote these bases 
\begin{equation}\label{eq:basisStatesGeneral}
	e \equiv \left\{ \left| e_{i}  \right\rangle: i=1,...,N \right\} ~~ {\rm and} ~~ 
	f \equiv \left\{ \left| f_{i} \right \rangle: i=1,...,N  \right\} \, .
\end{equation}
Here $N$ is the dimension of the photon states. The only condition imposed on $e$ and $f$ is that they form a basis.

\item Similarly, Alice should encode the information that she wants to transmit to Bob such that it cannot be deduced easily by Evan. To obtain a non-zero error rate in case of eavesdropping, she should either use non-orthogonal states (as in B92 \cite{B92}) or randomly switch between {\em at least} two sets of basis states (as in BB84 \cite{BB84}). For simplicity, we assume in the following that Alice prepares each photon randomly in one of the basis states of $e$ and $f$.

\item In order to establish strong correlations between Alice's input state and Bob's measurement outcome, Alice needs to reveal some information via classical communication. This information should be enough for Alice and Bob to obtain a shared secret key bit but not enough for the eavesdropper to deduce it. One possibility is that Alice announces which basis, $e$ or $f$, she used (as in BB84 \cite{BB84}). Another possibility is that Alice reveals the index $i$ of the respective basis state (as in Refs.~\cite{highDim,highDim2}). This does not reveal any information about the key as long as the states $|e_i \rangle$ and $|f_i \rangle$ with the same index $i$ encode different bits. In this paper we consider this second approach and show that it can guarantee relatively high error rates in the presence of an eavesdropper. 

\begin{table*}[t]
	\centering
		\begin{tabular}{|c||c|c|c|c|c||c|c|c|c|c|}
			\hline
				Index announced & \multicolumn{10}{|c|}{State measured by Bob} \\ \cline{2-11}
				by Alice & $|e_{1} \rangle $ & $ |e_{2} \rangle $ &	$ ... $ & $ ... $ & $ |e_{N} \rangle $ & 
				$ |f_{1} \rangle $ & $ |f_{2} \rangle $ &	$ ... $ & $ ... $ & $ |f_{N} \rangle $ \\
			\hline \hline
				$\textbf{1} $ &	$ \eta_{11} $ & $ \eta_{12} $ &	$ ... $ &	$ ... $ &	$ \eta_{1N} $ &	$ \mu_{11} $ &	$ \mu_{12} $ &	$ ... $ &	$ ... $ &	$ \mu_{1N} $ \\
			\hline
				$\textbf{2} $ &	$ \eta_{21} $ & $ \eta_{22} $ &	$ ... $ &	$ ... $ &	$ \eta_{2N} $ &	$ \mu_{21} $ &	$ \mu_{22} $ &	$ ... $ &	$ ... $ &	$ \mu_{2N} $ \\
			\hline
				$ ... $ &	$ ... $ & $ ... $ &	$ ... $ &	$ ... $ &	$ ... $ &	$ ... $ &	$ ... $ &	$ ... $ &	$ ... $ &	$ ... $ \\
			\hline
				$ ... $ &	$ ... $ & $ ... $ &	$ ... $ &	$ ... $ &	$ ... $ &	$ ... $ &	$ ... $ &	$ ... $ &	$ ... $ &	$ ... $ \\
			\hline
				$\textbf{N} $ &	$ \eta_{N1} $ & $ \eta_{N2} $ &	$ ... $ &	$ ... $ &	$ \eta_{NN} $ &	$ \mu_{N1} $ &	$ \mu_{N2} $ &	$ ... $ &	$ ... $ &	$ \mu_{NN} $ \\		
			\hline
		\end{tabular}
	\caption{Bob's interpretation of his measurement outcomes as a function of the index announced by Alice. The parameters $\eta_{ij}$ and $\mu_{ij}$ can assume the values ``0,'' ``1,'' or ``$\times$'' indicating whether a ``0," a ``1," or no bit is transmitted.}
	\label{tab:general}
\end{table*}
 
\item We now have a closer look at how Bob should interpret his measurement outcomes after Alice told him the index $i$ of her basis state. He can do this by using a table like Table~\ref{tab:general}. If Alice announces that she prepared the photon in a state with index $i$, Bob obtains ``$\eta_{ij}$'' when he measures $|e_j \rangle$ and he obtains ``$\mu_{ij}$'' when he measures $|f_j \rangle$. The parameters $\eta_{ij}$ and $\mu_{ij}$ in the table assume three different values,  ``0,'' ``1,'' or ``$\times$,'' depending on whether Bob obtains a ``0,'' a ``1,'' or no key bit is transmitted.

Suppose Alice sends a photon prepared in $|e_1 \rangle$ in order to transmit a ``0." This implies
\begin{equation} \label{eq2}
\eta_{11} = ``0" ~ {\rm or} ~ ``\!\!\times \!\!"
\end{equation} 
in order to avoid that Bob obtains a wrong key bit. The state $|f_1 \rangle$ has to encode a ``1" in this case. Otherwise, Evan knows that a ``0" is transmitted, when ``$i = 1$" is announced. Consequently, 
\begin{equation}
\mu_{11} = ``1" ~ {\rm or} ~ ``\!\!\times \!\!" \,. 
\end{equation}
Moreover, if Alice announces ``$i=1$" and Bob measures $|f_j \rangle$ with $j \neq 1$, then he knows for sure that she prepared her photon in $|e_1 \rangle$. Analogously, if Alice announces ``$i=1$" and Bob measures $|e_j \rangle$ with $j \neq 1$, then he knows that Alice prepared $|f_1 \rangle$. Alice and Bob should therefore choose
\begin{equation} \label{eq4}
\eta_{1j} = ``1" ~~ {\rm and} ~~ \mu_{1j} =``0" ~~ {\rm for~all} ~~ j \neq 1 \, .
\end{equation}
There is no need for Bob to ignore a measurement outcome with $j \neq 1$ since he always knows which key bit the photon encodes in this case.

\item It is indeed possible (c.f.~Eqs.~(\ref{eq2})-(\ref{eq4})) that Table \ref{tab:general} contains no crosses and that every detected photon transmits one bit of the cryptographic key. However, the minimum error rate in the case of eavesdropping is already known to be relatively low in this case \cite{highDim,highDim2}. We therefore assume here that Bob ignores the cases where his measured state has the index $i$ announced by Alice and choose 
\begin{equation} \label{55}
\eta_{ii} = \mu_{ii} = ``\!\! \times \!\! " \, . 
\end{equation}
This means a key bit can only be obtained when Bob's measurement basis is different from the one used by Alice to prepare the photon. One can easily check that Alice and Bob always obtain the same secret key bit under ideal conditions. 

\begin{table*}[t]
	\centering
		\begin{tabular}{|c||c|c|c|c||c|c|c|c|}
			\hline
				Index announced
				& \multicolumn{8}{|c|}{States measured by Bob} \\ \cline{2-9}
				by Alice & $|e_{1} \rangle $ & $ |e_{2} \rangle $ & $ |e_{3} \rangle $ & $ |e_{4} \rangle $ & 
				$ |f_{1} \rangle $ & $ |f_{2} \rangle $ &	$ |f_{3} \rangle $ & $ |f_{4} \rangle $ \\
			\hline \hline
				$\textbf{1} $ &	$ \times $ & $ 1 $ &	$ 1 $ &	$ 1 $ &	$ \times $ &	$ 0 $ &	$ 0 $ &	$ 0 $ \\
			\hline
				$\textbf{2} $ &	$ 1 $ & $ \times $ &	$ 1 $ &	$ 1 $ &	$ 0 $ &	$ \times $ &	$ 0 $ &	$ 0 $ \\
			\hline
				$\textbf{3} $ &	$ 1 $ & $ 1 $ &	$ \times $ &	$ 1 $ &	$ 0 $ &	$ 0 $ &	$ \times $ &	$ 0 $ \\
			\hline
				$\textbf{4} $ &	$ 1 $ & $ 1 $ &	$ 1 $ &	$ \times $ &	$ 0 $ &	$ 0 $ &	$ 0 $ &	$ \times $ \\		
			\hline
		\end{tabular}
	\caption{Bob's interpretation of his measurement outcomes as a function of the index announced by Alice for $N=4$. All states of $e$ encode a ``0,'' while all states of $f$ encode a ``1.'' A key bit is obtained whenever the index of Bob's state is different from Alice's index.} \label{tab:dim4}
\end{table*}

\item For symmetry reasons, Alice should have equally many states to encode a ``0" as she has to encode a ``1." Without restrictions we therefore assume in the following that all the states of $e$ encode a ``0" while all states of $f$ encode a ``1." This means, Bob obtains a ``1" whenever he measures a state $|e_j \rangle$ with $j$ different from Alice's index $i$. Analogously, he obtains a ``0" when he measures a state $|f_j \rangle$ with $j \neq i$. 
\end{enumerate} 

The final protocol is summarised in Table \ref{tab:dim4} for the case where Alice and Bob communicate using four-dimensional photon states. For arbitrary $N$, the entire scheme works as follows: 
\begin{enumerate}
	\item Alice generates a random key sequence of classical bits and randomly assigns each bit value a random index $i=1,\, 2, \, ..., \, N$.
	\item Alice then uses this sequence and sends single photons prepared accordingly either in $|e_i \rangle$ or $|f_i \rangle$ to Bob.
	\item Bob measures the state of every incoming photon, thereby randomly switching the measurement basis between $e$ and $f$. 
	\item Alice publicly announces the random sequence of indices $i$ used to establish the cryptographic key.
	\item Bob interprets his measurement outcomes accordingly, using, for example, Table \ref{tab:dim4}, if $N=4$. He obtains a key bit whenever his index is different from the index announced by Alice.
	\item Bob tells Alice which photon measurements have been successful and provide a bit of the secret key.
	\item Finally, Alice and Bob determine whether an eavesdropper introduced an error into their communication. Whenever this error rate is sufficiently small, Alice and Bob can assume that no eavesdropping has occurred.
\end{enumerate}

Notice that no conditions are imposed on $e$ and $f$ in this section other than them being bases. This gives us a lot of flexibility when maximising the security of the corresponding cryptographic protocol. In fact, the only difference between the above protocol and the direct communication scheme introduced in Refs.~\cite{highDim,highDim2} is that we avoid the assumption of $\langle e_i |f_i \rangle$ being zero. Alice and Bob therefore have to discard their measurement outcomes when both their states have the same index. As we shall see below, the payoff for the corresponding loss in efficiency is a relatively high error rate in the presence of an eavesdropper. 

\section{Eavesdropping} \label{eavesdropping}

In the quantum key distribution protocol proposed here, the index $i$ of the photon state in transmission is the only publicly announced information. This index does not reveal any information about the corresponding key bit since it equally likely encodes a ``0" as it encodes ``1." An eavesdropper can therefore only learn about the cryptographic key by performing quantum measurements on the transmitted photons. In the following we assume that Evan measures the state of every photon using a basis $g$ which is optimal for his purpose. Afterwards, he forwards his measurement outcome to Bob. This eavesdropping strategy is known as an {\em intercept-resend} attack. Although, it is not the most general eavesdropping attack, the corresponding error rate is a strong indication for the security of a cryptographic protocol. Our aim is to increase the minimum error rate introduced by an eavesdropper above the $25 \, \%$ of the original BB84 protocol \cite{BB84}. As already mentioned above,  there are different types of errors which Alice and Bob can consider.

\subsection{The index transmission error rate} \label{ITER}

In the following, $i$ denotes again the index of the photon state prepared by Alice and $j$ is the index of the basis vector measured by Bob. When Alice and Bob use different bases, $j$ can assume any value between 1 and $N$, even in the absence of any eavesdropping. However, when Alice and Bob use the same basis, $i$ and $j$ should be the same. To detect Evan's presence, Alice and Bob could therefore do the following: Alice should randomly select some photons which should not be used to obtain key bits. For these photons, she tells Bob exactly which states she prepared. Comparing this information with his own measurement outcomes, Bob can then easily calculate the index transmission error rate (ITER). 

An index transmission error occurs when a photon prepared in $|e_i \rangle$ ($|f_i \rangle$) is measured at Bob's end as $|e_j \rangle$ ($|f_j \rangle$) with $i \neq j$. Assuming that Alice prepares the $2N$ basis states of $e$ and $f$ with the same frequency and that Bob measures $e$ and $f$ with the same frequency, we find that the ITER of the proposed protocol equals
\begin{equation}\label{eq:eveProbg1e1}
P_{\rm ITER} = {1 \over 2N} \sum_{i=1}^N \sum_{k=1}^N \sum_{j \neq i} \Big[ \, | \langle e_i | g_k \rangle |^2 \cdot | \langle g_k | e_j \rangle |^2 + | \langle f_i | g_k \rangle |^2 \cdot | \langle g_k | f_j \rangle |^2 \, \Big] 
\end{equation} 
for a given set of bases $e$, $f$, and $g$. The states $|g_k \rangle$ denote Evan's possible measurement outcomes which are forwarded to Bob without alteration. In principle, Evan could change the state of the transmitted photon by guessing which state Alice prepared. However, this strategy is not expected to reduce the above error rate significantly. A non-zero overlap between the basis states of $e$ and $f$ ensures that there is always a certain probability to guess incorrectly.

\begin{table*}[t]
\centering
			\begin{tabular}{|c||c|c|c|}
			\hline
			Dimension $N$ & ITER & QBER \\
			\hline 
			{\bf 2} & 0.2500 & 0.5000  \\
			\hline
			{\bf 3} & 0.3340 & 0.4959  \\
			\hline
			{\bf 4} & 0.4003 & 0.4759  \\
			\hline
		      {\bf 5} & 0.4965 & 0.4389 \\
		      \hline 
		      {\bf 6} & 0.5843 & 0.4379  \\
             	\hline 
		\end{tabular}
	\caption{Minimum ITER and minimum QBER as a function of $N$ introduced by Evan in an intercept-resend eavesdropping attack when Alice and Bob use optimal bases $e$ and $f$. Both error rates are obtained from a numerical simulation which randomly generates $100\,000$ $f$'s and $100\, 000$ $g$'s and calculates all the corresponding values for $P_{\rm ITER}$ and $P_{\rm QBER}$ using Eqs.~(\ref{eq:eveProbg1e2}) and (\ref{eq:eve2}).}
	\label{tab:ComparisonoftheErrorRateswithdifferentdimensionalbasisstates}
\end{table*}

To simplify Eq.~(\ref{eq:eveProbg1e1}) we take the normalisation of all the relevant basis vectors into account which implies 
\begin{equation}\label{eq:basisStatesEve}
\sum_{j = 1}^N | \langle g_k | e_j \rangle |^2 \, = \, \sum_{j = 1}^N | \langle g_k | f_j \rangle |^2 \, 
= \, \sum_{k = 1}^N | \langle g_k | e_j \rangle |^2 \, = \, \sum_{k = 1}^N | \langle g_k | f_j \rangle |^2 \, = \, 1 \, .
\end{equation} 
Substituting these identities into Eq.~(\ref{eq:eveProbg1e1}), the expression for the ITER simplifies to 
\begin{eqnarray}\label{eq:eveProbg1e2}
P_{\rm ITER} &=& 1 - {1 \over 2N} \sum_{i=1}^N \sum_{k=1}^N \Big[ \, | \langle g_k | e_i \rangle |^4 + | \langle g_k | f_i \rangle |^4 \, \Big] \, . 
\end{eqnarray}
The same expression is obtained when calculating this error rate by subtracting the probability of not making an error under the condition that Alice and Bob use the same basis from unity. 

\begin{figure}
\begin{center}
{\includegraphics[width=0.60\textwidth]{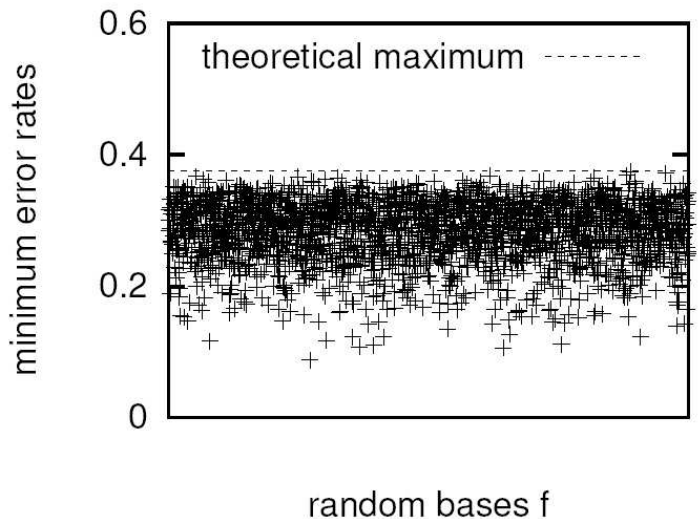}} 
\caption{Illustration of the numerical calculation of the minimum ITER introduced by Evan for  $N=4$ and $2 \, 500$ random choices of $f$. The crosses are the minima obtained after generating $10^6$ $g$'s and comparing the corresponding ITER's given by Eq.~(\ref{eq:eveProbg1e2}) for each $f$. The dotted line shows the theoretical maximum of $37.5 \, \%$ of these minima (cf.~Eq.~(\ref{concerror})).} \label{fig1}
\end{center}
\end{figure}

While Alice and Bob want the error rate in Eq.~(\ref{eq:eveProbg1e2}) to be as large as possible, Evan wants it to be as small as possible. Both parties, with Alice and Bob on one side and Evan on the other side, try to optimise the choice of the bases $e$, $f$, and $g$ accordingly. Table \ref{tab:ComparisonoftheErrorRateswithdifferentdimensionalbasisstates} shows the results of a numerical solution of this double-optimisation problem for different dimensions $N$. To obtain this table, the basis $e$ is kept fixed and a large number of bases $f$ is generated randomly. For each $f$ we then generate another large set of random bases $g$ and determine the minima of the corresponding error rates using Eq.~(\ref{eq:eveProbg1e2}). This is illustrated in Fig.~\ref{fig1} for the $N=4$ case. The ITER in Table \ref{tab:ComparisonoftheErrorRateswithdifferentdimensionalbasisstates} is the maximum of all the obtained minimum error rates. 

For $N=2$, we find that the minimum ITER introduced by an eavesdropper equals $25 \, \%$  when Alice and Bob use an optimal choice of $e$ and $f$, as in the original BB84 protocol \cite{BB84}. However, when Alice and Bob increase the dimension $N$ of their photon states, the minimum ITER increases. One can easily see that $50 \, \%$ constitutes an upper bound for the minimum ITER by considering the case where Evan measures the incoming photons either in the $e$ or in the $f$ basis. Using this eavesdropping strategy, the states of at least half of the transmitted photons remain unaffected. 

\subsection{The quantum bit error rate}
 
Alternatively, Alice and Bob can detect a potential eavesdropper by calculating the usual quantum bit error rate (QBER). To do so, both randomly select a certain number of control bits from the obtained key sequence and compare them openly. Notice that a key bit is obtained when the index $j$ of the state measured by Bob and the index $i$ of the state prepared by Alice are different. Bob interprets his measurement result correctly only when both states belong to a different bases. A quantum bit error hence occurs when Bob measures $|e_j \rangle$ ($|f_j \rangle$) while Alice prepared $|e_i \rangle$ ($|f_i \rangle$) with $i \neq j$. Using Eq.~(\ref{eq:eveProbg1e1}), the QBER for a given set of bases $e$, $f$, and $g$ can be written as
\begin{equation} \label{eq:eve1}
P_{\rm QBER} = {P_{\rm ITER} \over 2 P_{\rm IC}} \, ,
\end{equation} 
where the {\em index-change} (IC) probability $P_{\rm IC}$, 
\begin{eqnarray} \label{PIC}
P_{\rm IC} &=& {1 \over 4N} \sum_{i=1}^N \sum_{k=1}^N \sum_{j \neq i} \Big[ \, | \langle e_i | g_k \rangle |^2 \cdot | \langle g_k | e_j \rangle |^2 + | \langle f_i | g_k \rangle |^2 \cdot | \langle g_k | f_j \rangle |^2 \nonumber \\
&& + | \langle e_i | g_k \rangle |^2 \cdot | \langle g_k | f_j \rangle |^2 + | \langle f_i | g_k \rangle |^2 \cdot | \langle g_k | e_j \rangle |^2 \, \Big] \, ,
\end{eqnarray}  
is the probability that the index $j$ of Bob's state is different from the index $i$ of Alice's state. Using the identities in Eq.~(\ref{eq:basisStatesEve}), we find that the QBER in Eq.~(\ref{eq:eve1}) equals 
\begin{eqnarray}\label{eq:eve2}
P_{\rm QBER} &=& {2N - \sum_{i=1}^N \sum_{k=1}^N \Big[ \, | \langle e_i | g_k \rangle |^4 + | \langle f_i | g_k \rangle |^4 \, \Big] \over 4N - \sum_{i=1}^N \sum_{k=1}^N \Big[ \, | \langle e_i | g_k \rangle |^2 + | \langle f_i | g_k \rangle |^2 \, \Big]^2} \, .
\end{eqnarray}  
The third column in Table 3 shows the minimum QBER introduced by an eavesdropper, when Alice and Bob use optimal bases $e$ and $f$, for different dimensions $N$. As in Section \ref{ITER}, these probabilities have been obtained by comparing probabilities for a large set of randomly generated bases $f$ and $g$.

Even for $N=2$, the minimum QBER can be as high as $50 \, \%$. A more detailed analysis of the corresponding protocol shows that Alice and Bob can realise this scenario by choosing $e$ and $f$ almost identical, independent of Evan's choice of measurement basis $g$. However, the price they pay for this very high QBER is a steep drop in the efficiency of their quantum key distribution. In the extreme case, where $|e_1 \rangle \equiv |f_1 \rangle$ and $|e_2 \rangle \equiv |f_2 \rangle$, it becomes impossible to generate secret key bits, since Alice's and Bob's state always have the same index $i$, at least in the absence of any eavesdropping. In the following, we assume therefore  that Alice and Bob use the ITER in order to detect the presence of an eavesdropper. 

\section{Optimal choices of $e$, $f$, and $g$} \label{concrete}

We now address the question of how Alice and Bob can take advantage of the high ITER's shown in Table \ref{tab:ComparisonoftheErrorRateswithdifferentdimensionalbasisstates} by having a look at possible realisations of the proposed quantum key distribution protocol. First, we consider the $N=2$ case which suggests an optimal strategy for Alice and Bob in higher dimensions. Moreover, we discuss in this section what Evan can do to best cloud his presence.

\subsection{The $N=2$ case} \label{sec:x}

\begin{figure}[t]
\begin{center}
\includegraphics[width=0.40\textwidth]{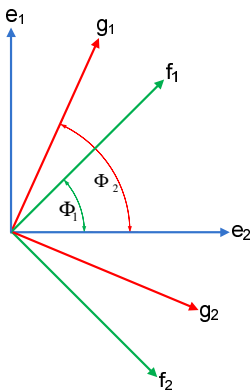}
\caption{Basis vectors used by Alice, Bob, and Evan in the $N=2$ protocol. To maximise the minimum error rate introduced by Evan in case of an intercept-resend attack, Alice and Bob should choose $\phi_1 = {1 \over 4} \pi$. This yields an error rate of $25 \, \%$, independent of Evan's choice of $\phi_2$.} \label{fig2}
\end{center}
\end{figure}

The problems which Alice, Bob, and Evan have to solve in the $N=2$ case in order to optimise their strategies are exactly the same as in BB84 \cite{BB84}. Suppose Alice and Bob choose (cf.~Fig.~\ref{fig2})
\begin{eqnarray}\label{eq:optimalValuesE}
|e_1 \rangle  = \left( \begin{array}{r} 1 \\ 0 \end{array} \right) , ~
|e_2 \rangle  = \left( \begin{array}{r} 0 \\ 1 \end{array} \right) , ~
\end{eqnarray}
while $|f_1 \rangle $ and $|f_2 \rangle $ are, without restrictions, given by
\begin{eqnarray}\label{eq:optimalValuesF}
|f_1 \rangle  = \left( \begin{array}{r} \cos \phi_1 \\ \sin \phi_1 \end{array} \right) , ~
|f_2 \rangle  = \left( \begin{array}{r} - \sin \phi_1 \\ \cos \phi_1 \end{array} \right) . 
\end{eqnarray}
Moreover, we write the states of Evan's optimal measurement basis as
\begin{eqnarray}\label{eq:optimalValuesF}
|g_1 \rangle  = \left( \begin{array}{r} \cos \phi_2 \\ \sin \phi_2 \end{array} \right) , ~
|g_2 \rangle  = \left( \begin{array}{r} - \sin \phi_2 \\ \cos \phi_2 \end{array} \right) . 
\end{eqnarray}
Substituting these states into Eq.~(\ref{eq:eveProbg1e2}), we find
\begin{eqnarray}
P_{\rm ITER} &=& {1 \over 4} \left[ \, \sin^2 \big(2 (\phi_1 - \phi_2) \big) + \sin^2 (2 \phi_2) \, \right] \, .
\end{eqnarray}
In order to maximise the minimum of this probability with respect to $\phi_2$, Alice and Bob should choose $\phi_1 = {1 \over 4} \pi$. In this case, $\sin^2 \big(2 (\phi_1 - \phi_2) \big)$ becomes the same as $\cos^2 (2 \phi_2)$. This error rate equals $25 \, \%$ independent of Evan's choice of $\phi_2$. For the eavesdropper, every possible strategy is hence an optimal one.

In other words, for $N=2$, Alice and Bob's optimal choice for $e$ and $f$ are two {\em mutually unbiased} bases \cite{Wootters}. This means, upon measurement, a photon prepared in any of the basis states of $e$ is found with equal probability in any of the basis states of $f$ and vice versa. As shown in Fig.~\ref{fig2}, $e$ and $f$ should be as far away from each other as possible. A straightforward generalisation of this result to higher dimensions suggests that Alice and Bob should always use two mutually unbiased bases $e$ and $f$.  

\subsection{The $N=4$ case} \label{sec:y}

\begin{figure}[t]
\begin{center}
{\includegraphics[width=0.60\textwidth]{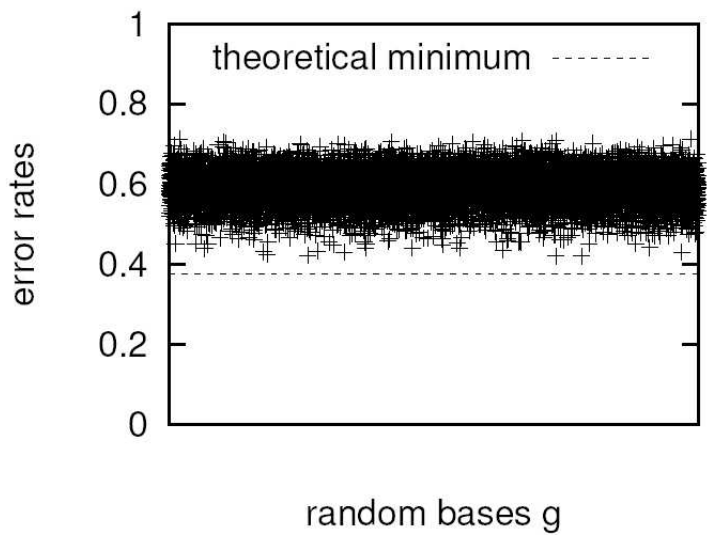}} 
\caption{Numerical calculation of the error rates introduced by Evan for $10^5$ randomly generated four-dimensional $g$'s while the states of $e$ and $f$ are as in Eqs.~(\ref{e's}) and (\ref{f's}). We see that these rates are always above their theoretical minimum (cf.~Eq.~(\ref{concerror})) of $37.5 \, \%$ (dotted line) which is obtained when Evan measures either $e$ or $f$.} \label{fig4}
\end{center}
\end{figure}

Let us now have a closer look at the case where Alice and Bob communicate with four-dimensional photon states. To obtain two mutually unbiased bases, they could choose for example
\begin{eqnarray} \label{e's}
\hspace*{-1cm} |e_1 \rangle  = \left( \begin{array}{r} 1 \\ 0 \\ 0 \\ 0 \end{array} \right) , ~
|e_2 \rangle  = \left( \begin{array}{r} 0 \\ 1 \\ 0 \\ 0 \end{array} \right) , ~ 
|e_3 \rangle  = \left( \begin{array}{r} 0 \\ 0 \\ 1 \\  0 \end{array} \right) , ~ 
|e_4 \rangle  = \left( \begin{array}{r} 0 \\ 0 \\ 0 \\ 1 \end{array} \right) ,
\end{eqnarray}
while the states of $f$ are given by
\begin{eqnarray} \label{f's}
\hspace*{-1cm} |f_1 \rangle  = {1 \over 2} \left( \begin{array}{r} 1 \\ 1 \\ -1 \\ -1 \end{array} \right) , ~
|f_2 \rangle  = {1 \over 2} \left( \begin{array}{r} -1 \\ 1 \\ 1 \\ -1 \end{array} \right) , ~
|f_3 \rangle  = {1 \over 2} \left( \begin{array}{r} 1 \\ -1 \\ 1 \\ -1 \end{array} \right) , ~
|f_4 \rangle  = {1 \over 2} \left( \begin{array}{r} 1 \\ 1 \\1 \\ 1 \end{array} \right) . \nonumber \\
\end{eqnarray}
To find an optimal intercept-resend strategy for the eavesdropper, we assume that he measures states which lie on a line between the closest states of the $e$ and the $f$ basis. More concretely, we choose
\begin{eqnarray} \label{gi}
|g_i \rangle  = {\cos \alpha \, |e_i \rangle + \sin \alpha \, |f_i \rangle  \over \left( 1 + {1 \over 2} \sin (2 \alpha) \right)^{1/2}} \, .
\end{eqnarray}
One can easily check that the $|g_i \rangle$'s are normalised and pairwise orthogonal. This applies since the indices of the basis states in Eqs.~(\ref{e's}) and (\ref{f's}) have been chosen accordingly. Using Eq.~(\ref{eq:eveProbg1e2}), we find that the ITER introduced by Evan now equals 
\begin{eqnarray} \label{error}
P_{\rm ITER} &=& {3 \over 8} \left[ 1 + { \sin^2 (2 \alpha) \over \left( 2 + \sin (2 \alpha) \right)^2} \right] \, .
\end{eqnarray}
Since the second term in the brackets is always positive, one can easily see that $P_{\rm ITER} \ge 37.5 \, \%$. Indeed, the best strategy for Evan is to choose $\sin (2 \alpha) = 0$. This means, Evan should measure either $e$ or $f$. 

Fig.~\ref{fig4} shows the error rate introduced by Evan for the above choice of $e$ and $f$ and for a large set of randomly generated $g$'s with $N=4$. It confirms that $P_{\rm ITER}$ is always above $37.5 \, \%$, if Alice and Bob use two mutually unbiased bases. This applies even when no assumption on the form of the states of $g$ is made, as we do for our analytical calculations in Eq.~(\ref{gi}). A comparison of $P_{\rm ITER} = 37.5 \, \%$ with the result in Table \ref{tab:ComparisonoftheErrorRateswithdifferentdimensionalbasisstates} for $N=4$ confirms that this error rate corresponds to (or is at least very close to) an optimal strategy of Alice, Bob, and Evan. Both results agree within the error limits of the underlying numerical calculation.

\subsection{The general case}

Let us now have a look at the optimal choice of $e$, $f$, and $g$ for the general case where Alice and Bob use $N$-dimensional photon states. As suggested at the end of Section \ref{sec:x}, we assume that Alice and Bob use two mutually unbiased bases. More concretely, we assume that their states are given by 
\begin{eqnarray} \label{last}
e_{ij} = \delta_{ij} ~~ {\rm and} ~~ f_{ij} = {1\over \sqrt{N}} \, \omega_N^{(i-1)(j-1)} ~~ {\rm with} ~~ \omega_N = \exp \left({2 {\rm i} \pi \over N} \right) \, ,
\end{eqnarray}
$|e_i \rangle \equiv (e_{i1},e_{i2}, ... , e_{iN})^{\rm T}$ and $|f_i \rangle \equiv (f_{i1},f_{i2}, ... , f_{iN})^{\rm T}$. One can easily check that $e$ and $f$ are orthonormal and mutually unbiased. We then generate a large set of random $g$'s and calculate the corresponding error rates $P_{\rm ITER}$ using Eq.~(\ref{eq:eveProbg1e2}). Table \ref{tab:more} shows the maxima of these rates as a function of $N$. The given error rates hence correspond to Evan's optimal intercept-resend eavesdropping strategy. 

\begin{table*}[t]
\centering
			\begin{tabular}{|c||c|c|c|c|}
			\hline
			Dimension $N$ 
			& $\begin{array}{c} {\rm ITER} \\[-0.1cm] {\rm (numerics)} \end{array}$ 
			& $\begin{array}{c} {\rm ITER} \\[-0.1cm] {\rm (analytics)} \end{array}$ \\
			\hline 
			{\bf 2} & 0.2500 & 0.2500 \\
			\hline
			{\bf 3} & 0.3333 & 0.3333 \\
			\hline
			{\bf 4} & 0.3794 & 0.3750 \\
			\hline
			{\bf 5} & 0.4529 & 0.4000 \\
			\hline			
			{\bf 6} & 0.5348 & 0.4167 \\
			\hline
			{\bf 7} & 0.6214 & 0.4286 \\
			\hline
			{\bf 8} & 0.6707 & 0.4375 \\
			\hline 
		\end{tabular}
	\caption{Minimum ITER introduced by an eavesdropper in case of an intercept-resend attack when Alice and Bob use the mutually unbiased bases $e$ and $f$ of Eq.~(\ref{last}). The second column is the result of a numerical simulation which generates $5 \, 000 \, 000$ $g$'s, calculates the respective error rates using Eq.~(\ref{eq:eveProbg1e2}), and determines their minimum. Third column shows the theoretical values given by Eq.~(\ref{concerror}).} \label{tab:more}
\end{table*}

A comparison of the ITER's in Table \ref{tab:more} with the ITER's in Table \ref{tab:ComparisonoftheErrorRateswithdifferentdimensionalbasisstates} confirms that using two mutually unbiased bases $e$ and $f$ is (at least close to) an optimal strategy for Alice and Bob. For $N=2$, we find again that the minimum error rate introduced by Evan equals $25 \, \%$. For $N=3$ this rate equals $33 \, \%$, and for higher-dimensional photon states, the values in the third column of Table \ref{tab:more} approach their predicted maximum of $50 \, \%$ (cf.~Section \ref{eavesdropping}). The second column has been obtained from a numerical optimisation of Evan's strategy. Since it is a relatively hard computational problem to find the best eavesdropping measurement basis $g$ numerically, the errors in this column are relatively large, especially for large $N$. 

Let us now have a closer look at the best intercept-resend eavesdropping strategy for Evan. The discussion of the $N=4$ case in Section \ref{sec:y} suggests that Evan should measure either $e$ or $f$ in order to minimise the bit transmission error rate. If $e$ and $f$ are mutually unbiased, then the probability of detecting a photon in $|e_i \rangle$ equals $1/N$ when Alice prepares an $f$-state. Analogously, the probability of detecting a photon in $|f_i \rangle$ equals $1/N$ when Alice prepares an $e$-state. Substituting this into Eq.~(\ref{eq:eveProbg1e2}) yields 
\begin{eqnarray} \label{concerror}
P_{\rm ITER} &=& {N-1 \over 2N} \, . 
\end{eqnarray}
For completeness we mention that the corresponding QBER (cf.~Eq.~(\ref{eq:eve2})) equals $33 \, \%$ independent of $N$. A comparison with a numerical evaluation of the ITER confirms that measuring either $e$ or $f$ and forwarding the respective measurement outcome to Bob is indeed an optimal (or at least a close to optimal) strategy for Evan, if Alice and Bob test his presence by calculating this error rate.  

Let us conclude this subsection by commenting on the efficiency of the described quantum key distribution scheme. As in BB84, Alice and Bob randomly switch between two sets of basis states. Here a key bit can only be obtained when both use a {\em different} basis. Moreover, the index of the state measured by Bob should be different from the index of Alice's state. The probability for this to happen and hence the mean number of bits per transmitted photon equals 
\begin{eqnarray} \label{success}
P_{\rm success} &=& {N-1 \over 2N} \, . 
\end{eqnarray}
This expression is exactly the same as the ITER in Eq.~(\ref{concerror}).

\subsection{Possible implementation}

In order to implement the above protocol, Alice needs a single photon source. As in BB84, the photon can come from a parametric down conversion crystal, a very weak laser pulse, or an on-demand single-photon source. Using path encoding, the states of $f$ can be prepared easily with the help of a  Bell multiport beam splitter. Such a beam splitter may consist of a network of beam splitters and phase plates \cite{Reck94,Zukowski97} which have to be interferometrically stable. It can also be made by splicing $N$ optical fibers \cite{Pryde03}. Spliced fibre constructions are commercially available and can include between three and thirty input and output ports.

The main feature of a symmetric $N \times N$ Bell multiport beam splitter is that a photon entering any of its input ports is redirected with equal likelihood to any of its $N$ possible output ports. One way for Alice to prepare the state $|e_i \rangle$ in Eq.~(\ref{last}) is to bypass the beam splitter and to send a single photon directly to output port $i$. In this case, preparing the state $|f_i \rangle$ in Eq.~(\ref{last}) only requires to send a single photon into input port $i$ \cite{Lim}. Bob can use the same setup as Alice to decode the key bit. To measure $f$, he should send the incoming photon first through a Bell multiport beam splitter and then detect it in one of the $N$ output ports. To measure $e$, he can simply bypass this step. During the transmission, the path encoding should be switched to time-bin encoding which promises a better protection of the photons against decoherence \cite{timebin}.

\section{Conclusions} \label{conclusions}

In this paper we propose a quantum key distribution protocol where Alice and Bob use higher-dimensional photon states. The scheme does not encode information in a higher-dimensional alphabet \cite{Bechmann,Bechmann2,Cerf,Cerf2,Bruss,Sych,Walborn} and is not a straightforward generalisation of the original BB84 protocol \cite{BB84}. Instead, Alice and Bob use two bases $e$ and $f$ with all $e$-states encoding a ``0" and all $f$-states encoding a ``1." Under ideal conditions, a key bit is obtained when Alice and Bob use different bases. In Section \ref{design}, no conditions are imposed on the states of $e$ and $f$ which gives us the flexibility to maximise the error rate introduced by an eavesdropper in case of an intercept-resend attack. This is not the only possible eavesdropping strategy but security against this is a strong indication for the general security of a cryptographic protocol. 

In Section \ref{eavesdropping}, we generate large sets of random basis states and determine the minimum index transmission error rate (ITER) introduced by an eavesdropper numerically. For $N=2$, this error rate equals the $25 \, \%$ quantum bit error rate (QBER) of the BB84 protocol \cite{BB84}. However, the minimum ITER rapidly approaches $50 \, \%$ in higher dimensions (cf.~Table \ref{tab:ComparisonoftheErrorRateswithdifferentdimensionalbasisstates}). A detailed analysis of the $N=2$ and the $N=4$ case suggests that Alice and Bob should use two mutually unbiased basis $e$ and $f$. The best an eavesdropper can do to hide his presence is to measure the transmitted photons either in $e$ or $f$. This hypothesis is consistent with the numerical results in Section \ref{concrete} (cf.~Table \ref{tab:more}). Finally we point out that the proposed quantum key distribution protocol can be implemented for example with the help of a symmetric Bell multiport beam splitter \cite{Lim} and switching from path to time-bin encoding during the transmission. The mean number of key bits per transmitted photon turns out to be of exactly the same as the minimum error rate introduced by Evan (cf.~Eq.~(\ref{success})). 

In Section \ref{eavesdropping} we point out that it is in principle possible to obtain a minimum QBER close to $50 \, \%$, even for $N=2$. This requires Alice and Bob to use two bases $e$ and $f$ which are almost identical. Unfortunately, this strategy corresponds to a very low efficiency of the proposed cryptographic protocol. For $e \equiv f$, the key transmission rate drops to zero. Analytic expressions for the QBER and the efficiency of the bit transmission in the presence of an eavesdropper for a given set of bases can be found in Eqs.~(\ref{PIC}) and (\ref{eq:eve2}).  

\ack

We thank H. Zbinden for helpful comments. MMK thanks J. Xu for encouraging discussions and acknowledges funding from the NED University of Engineering $\&$ Technology, Karachi, Pakistan. A.B. thanks the project students S. So and P. Woodward for their initial participation in this project and acknowledges a James Ellis University Research Fellowship from the Royal Society and the GCHQ. This project was supported in part by the UK Engineering and Physical Sciences Research Council through the QIP IRC and the EU Research and Training Network EMALI.

\end{document}